# Waku: A Family of Modular P2P Protocols For Secure & Censorship-Resistant Communication

Oskar Thorén*[†], Sanaz Taheri-Boshrooyeh*[†], Hanno Cornelius*[†]
*Vac Research and Development, [†]Status Research and Development, Singapore
oskar@vac.dev, sanaz@vac.dev, hanno@vac.dev

*Abstract*—Waku is a family of modular protocols that enable secure, censorship-resistant, and anonymous peer-to-peer communication. Waku protocols provide capabilities that make them suitable to run in resource-restricted environments e.g., mobile devices and web browsers. Such capabilities include (i) retrieving historical messaging for mostly-offline devices (ii) adaptive nodes; allowing for heterogeneous nodes to contribute to the network (iii) preserving bandwidth usage for resource-restricted devices, (iv) minimizing connectivity requirements for devices with a limited connection, and (v) enabling efficient, private, economic spam protection for heterogeneous nodes. Waku's modular design and resource-efficient protocols make it superior to its predecessor i.e., Whisper. In this paper, we give an overview of the Waku protocols stack, its architecture, and protocols interaction along with a sample demo scenario on configuring and running a Waku node using nwaku i.e., Waku client written in Nim.

## I. INTRODUCTION

Waku is a family of modular protocols designed to enable privacy-preserving peer-to-peer communication for decentralized applications. It is the successor of Whisper [1], the messaging layer of the Ethereum blockchain P2P protocol suite, and outperforms it in scalability and resource efficiency. Below we highlight Waku's features[1]. At the time of this article, Waku is deployed by *Status* and *WalletConnect v2*.

**Peer-to-peer & Censorship-resistant**: Waku protocols follow a p2p design where an intended functionality is realized by the resources supplied by independently run peers. Being p2p makes Waku protocols resistant to censorship and a single point of failure also allows utilization of shared infrastructure.

**Privacy-Preserving**: Unlike centralized messaging systems, Waku does not rely on peers' Personally Identifiable Information (PII) e.g., phone number or email address. Also, protocol messages are metadata/PII protected to enable user anonymity and various forms of unlinkability [2].

**Useful for generalized messaging**: Waku is designed to work in various messaging scenarios including human-to-human e.g., a messenger, or machine-to-machine e.g., state channels.

**Modularity**: Waku protocols are modular hence nodes can adaptively choose protocols to support. We call this concept *adaptive nodes* explained in 30/ADAPTIVE-NODES RFC [3]. This allows applications to make trade-offs based on the properties they and their users value, e.g., resource usage vs metadata protection; providing useful services to the network vs being just a consumer.

**Runs anywhere**: Waku protocols can be run in environments with limited resources e.g., bandwidth, CPU, memory, disk, and battery, and connectivity (i.e., not being publicly connectable, only being intermittently connected, and mostly-offline).

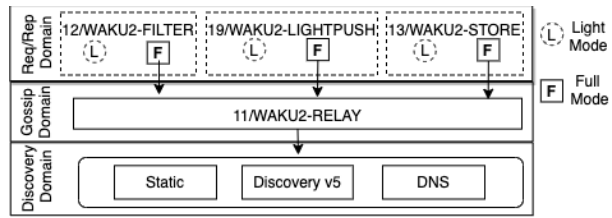

Figure 1: Stack of Waku protocols. Arrows represent protocols dependency. Dashed-line rectangles indicate protocols suitable for resource-limited nodes.

## II. SOFTWARE ARCHITECTURE

We present four core protocols of the Waku stack as illustrated in Figure 1. We note that this is not an exhaustive list. Protocols are referred to by their RFC IDs as in their reference specifications [3] and are mostly implemented over libp2p and associated with libp2p protocol identifiers. Protocols fall into one of the three separate network interaction domains i.e., (a) Gossip domain, (b) Discovery domain, and (c) Request/Reply (Req/Rep) domain as explained next.

**Waku Message:** Waku protocols are used to communicate a data unit called a Waku Message i.e., 14/WAKU2-MESSAGE [3]. Each message is associated with a content topic used for content filtering. The content topic has no impact on the message routing, but enables fine-grain queries in Req/Reply protocols and can be used to specify an application domain.

**Gossipsub Domain**: Waku uses a publisher-subscriber messaging model to disseminate Waku messages. This is done through the 11/WAKU2-RELAY protocol which is a thin layer on top of libp2p GossipSub [4]. It is a Gossip-based routing algorithm where messages are published to topics and delivered by gossiping to peers subscribed to that pubsub topic. Waku also features an experimental privacy-preserving economic and efficient spam protection mechanism on top of 11/WAKU2-RELAY called 17/WAKU2-RLNRELAY [3].

**Discovery Domain** Waku supports multiple discovery methods by which nodes become aware of each other's existence. That includes DNS-based discovery as per EIP-1459 [5] by which a new node can retrieve an authenticated, updateable list of peers to bootstrap connection to the network. This method merits high scalability and resource efficiency, though, is centralized hence prone to attacks e.g., censorship. It may be used in conjunction with other ambient peer discovery methods e.g., Node Discovery v5 [6] which is a DHT-based mechanism and, unlike the DNS-based approach, is decentralized. Nevertheless, its high resource usage e.g., CPU and memory must be carefully considered when used for resource-limited devices. Waku also supports the basic form of discovery where a set of static nodes may be specified.

---

[1]This is the first paper aimed at turning the technical aspects of the Waku protocol stack into an academic-grade publication. Hence, most of the references are links to the available production-grade industry-standard specifications i.e., RFCs.

**Request/Reply Domain**: Waku supplies a set of Req/Rep protocols primarily used to get Waku to run in resource-restricted environments e.g., with limited bandwidth or connectivity. Req/Rep protocols have two modes 1) full mode; peer can act as both a requester and a responder 2) light mode; peer is only a requester. The prerequisite of running in full mode is to have support for 11/WAKU2-RELAY protocol.

- 13/WAKU2-STORE provides support for historical messages where it enables querying of Waku messages received through 11/WAKU2-RELAY protocol and stored by other nodes. This is especially useful for mostly offline devices. Nodes running this protocol in full mode are expected to have high availability and ample storage. Waku also provides an experimental fault-tolerant addition to the store protocol called 21/WAKU2-FT-STORE [3] that relaxes the high availability requirement.
- 12/WAKU2-FILTER is a lightweight version of 11/WAKU2-RELAY designed for bandwidth restricted devices in which nodes running in light mode can subscribe to nodes running in full mode to only receive Waku messages they desire e.g., based on the pubsub topic or content topic.
- 19/WAKU2-LIGHTPUSH is used for nodes with short connection windows and limited bandwidth to publish messages. A receiver of light push message publishes the encapsulated Waku message on the requested pubsub topic.

A sample interaction of Waku protocols is visualized in Figure 2. There are three nodes denoted by *A-C*. The PubSub topic `pub1` is used for routing and *B* is subscribed to that topic. *B* also supports the 13/WAKU2-STORE protocol. *A* creates `msg1` with content topic `content1` and publishes it on `pub1`. As *B* is subscribed to that topic, it picks up the message and saves it for possible later retrieval by other nodes. At a later time, *C* comes online and requests messages matching `pub1` and `content1` from *B*. *B* responds with messages meeting this, and possibly other, criteria.

### III. SAMPLE DEMO SCENARIO

*nwaku* [7] is an open-source and reference implementation of the Waku client written in Nim programming language. There are also other implementations of Waku available that support a subset of Waku protocols, namely, js-waku [8] and go-waku [9], written in Javascript and Go, respectively.

Listing 1 illustrates a sample configuration of a Waku node using *nwaku* client via a set of command-line flags. `wakunode2` is the binary executable of *nwaku* which can be built as instructed in [7]. The node is configured to statically connect to a relay node whose libp2p multi-address [10] is passed in `staticnode` option. The `relay` flag enables 11/WAKU2-RELAY protocol over the pubsub topic supplied by `topics` flag which is `/waku/2/default-waku/proto`. Similarly, the 13/WAKU2-STORE protocol is enabled by passing the `store` option. The boolean flag `persist-messages` is used to distinguish between the full-mode and light mode of the store protocol (`true` indicates full mode). The `store-capacity` specifies a limit on the number of persisted Waku messages. The `filter` and `lightpush` flags are used to mount the 12/WAKU2-FILTER protocol and 19/WAKU2-LIGHTPUSH in full mode. A Waku node may optionally expose a set of JSON RPC APIs using the `rpc` flag depending on its supported protocols. More details on these APIs are given in 16/WAKU2-RPC RFC [3]. The corresponding IP address and port can be set using `rpcAddress` and `rpc-port` options.

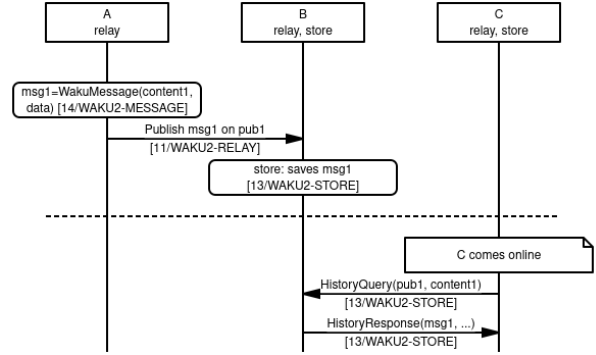

Figure 2: Simplified protocol interaction in Waku.

```
wakunode2
--staticnode:/ip4/134.209.139.210/tcp/30303/p2p/16Ui
u2HAmPLe7Mzm8TsYUubgCAW1aJoeFScxrLj8ppHFivPo97bUZ
--relay:true  --topics:/waku/2/default-waku/proto
--store:true  --persist-messages:true
--store-capacity:1000
--filter:true  --lightpush:true
--rpc:true --rpcAddress:127.0.0.1 --rpc-port:8545
```

Listing 1: A sample configuration of a Waku node

### REFERENCES

[1] "Whisper specification," https://eips.ethereum.org/EIPS/eip-627, accessed: 2022-03.
[2] "Security analysis of waku2-relay," https://rfc.vac.dev/spec/11/security-analysis, accessed: 2022-03.
[3] "Waku specifications," https://rfc.vac.dev, accessed: 2022-03.
[4] D. Vyzovitis and Y. Psaras, "Gossipsub: A secure pubsub protocol for unstructured, decentralised p2p overlays," 2019.
[5] "Node discovery via dns," https://eips.ethereum.org/EIPS/eip-1459, accessed: 2022-03.
[6] "Node discovery protocol version 5.1," https://github.com/ethereum/devp2p/blob/master/discv5/discv5.md, accessed: 2022-03.
[7] "Reference implementation of waku in nim," https://github.com/status-im/nim-waku, accessed: 2022-03.
[8] "Javascript implementation of waku," https://github.com/status-im/js-waku, accessed: 2022-03.
[9] "Go implementation of waku," https://github.com/status-im/go-waku, accessed: 2022-03.
[10] "Libp2p multi-address documentation," https://docs.libp2p.io/concepts/addressing/, accessed: 2022-03.